\documentclass[aps,prl,twocolumn,groupedaddress]{revtex4}

\usepackage{subfigure}

\usepackage{ifpdf}
\ifpdf
\usepackage[pdftex]{graphicx}
\else
\usepackage{graphicx}
\fi
\usepackage{hyperref}

\begin{document}

\ifpdf
\DeclareGraphicsExtensions{.pdf, .jpg}
\else
\DeclareGraphicsExtensions{.eps, .jpg}
\fi

\def\hslash{\hbar}
\def\imag{i}
\def\grad{\vec{\nabla}}
\def\div{\vec{\nabla}\cdot}
\def\curl{\vec{\nabla}\times}
\def\DDt{\frac{d}{dt}}
\def\ddt{\frac{\partial}{\partial t}}
\def\ddx{\frac{\partial}{\partial x}}
\def\ddy{\frac{\partial}{\partial y}}
\def\lap{\nabla^{2}}
\def\divv{\vec{\nabla}\cdot\vec{v}}
\def\gradS{\vec{\nabla}S}
\def\vvec{\vec{v}}
\def\wc{\omega_{c}}
\def\<{\langle}
\def\>{\rangle}
\def\Tr{{\rm Tr}}
\def\Csch{{\rm csch}}
\def\Coth{{\rm coth}}
\def\Tanh{{\rm tanh}}
\def\g2{g^{(2)}}

%\preprint{}

\title{On the internal photorelaxation mechanism of DNA}
\author{Arkadiusz Czader}
%%\email[]{Your e-mail address}
\affiliation{Department of Chemistry, University of Houston, Houston, Texas 77204, USA}
\author{Eric R. Bittner}
\affiliation{Department of Chemistry, University of Houston, Houston, Texas 77204, USA}

\date{\today}

\begin{abstract}
We propose a model for the photo-deactivation mechanism for DNA based upon accurate quantum chemical and molecular dynamical evaluations of model Watson/Crick nucleoside pairs and stacked pairs. Our results corroborate recent ultrafast experimental studies on DNA oligonucleotides and suggest that following photo-excitation to a local $\pi-\pi^*$ state, the excitation is rapidly delocalized over several (3-4) bases on an ultrafast time-scale.  However, this delocalized state is unstable with respect to the motions of the protons involved in hydrogen-bonding between Watson/Crick pairs and rapidly re-localizes to a charge-transfer state on a longer time-scale ranging from 10 to 100 ps. This state, too, is unstable and relaxes via a conical intersection with the ground state near the geometry of the enol- and imino-tautomeric form. We suggest that this  internal deactivation mechanism is responsible for the intrinsic photostability of DNA.  
\end{abstract}
\maketitle

The mutagenic effect of the UV irradiation on the genetic code has been a topic of interest in the last few decades. As a result of the absorption of the short-wavelength component of UV light ($\lambda < 290$ nm) by the stratospheric ozone layer the spectrum of sunlight at the sea level barely overlap the low energy absorption tail of the nucleic acid bases: adenine (A), thymine (T), guanine (G), and cytosine (C). However, the depletion of the ozone layer in the recent years raises the concern of increasing exposure to the UV radiation. Existing reports indicate that for every 1\% decrease in the ozone layer there could be 4\% increase in the incidence of skin cancer. Photoexciation of DNA in the UVB and UVA region gives rise to photolesions, primarily in the form of cyclobutane pyrimidine dimers (CPD) and the pyrimidine(6-4)pyrimidone photoadducts. \cite{Mouret:2006,Gruijl:2000} The presence of these lesions are associated with a number  of nonmelanomic skin cancers.\cite{Cadeta:2005} 
The adverse effect of these photolesions is largely mitigated via enzymatic action involving DNA photolyase;  however, such  secondary mechanisms are both error prone and energetically costly and most certainly developed later in the evolutionary process. Experimental evidence also suggests that photoexcitations in DNA can be rapidly quenched through internal conversion
and dissipated as heat.  \cite{harrison:7001,Sinha:2002}

Until development of the femtosecond laser spectroscopy,  attempts to study the dynamics of these excited states were thwarted by the ultrafast character of the internal conversion process. In the last decade the combined effort of the researchers employing femtosecond spectroscopy, molecular beam techniques, and quantum chemistry methods shed light on the excited-state dynamics of the nucleic acids. However the precursor state which initiates the photochemical reactions damaging the genome remains elusive. Ultrafast spectroscopic studies by Fiebig {\em et al.} \cite{Rist:2002,Buchvarov_2007} indicate that following photo-excitation, the exciton rapidly delocalizes over a number of stacked bases. A similar explanation of the long-lived excited states was offered by Markovitsi and coworkers. They interpret the results of the fluorescence upconversion and time-correlated single photon counting  experiments in the frame of the exciton theory. In this model the large number of neutral excitons states are formed over several bases located both on the same strand and on opposite strands with subsequent relaxation of the exciton. In this approach both base stacking and base pairing determine the excited state  dynamics.\cite{Bouvier2002vu,bouvier:13512,Markovitsi:17130,Emanuele:2005a,emanuele:16109}

\begin{figure}[t]
\center{
\includegraphics[width=1.0\columnwidth]{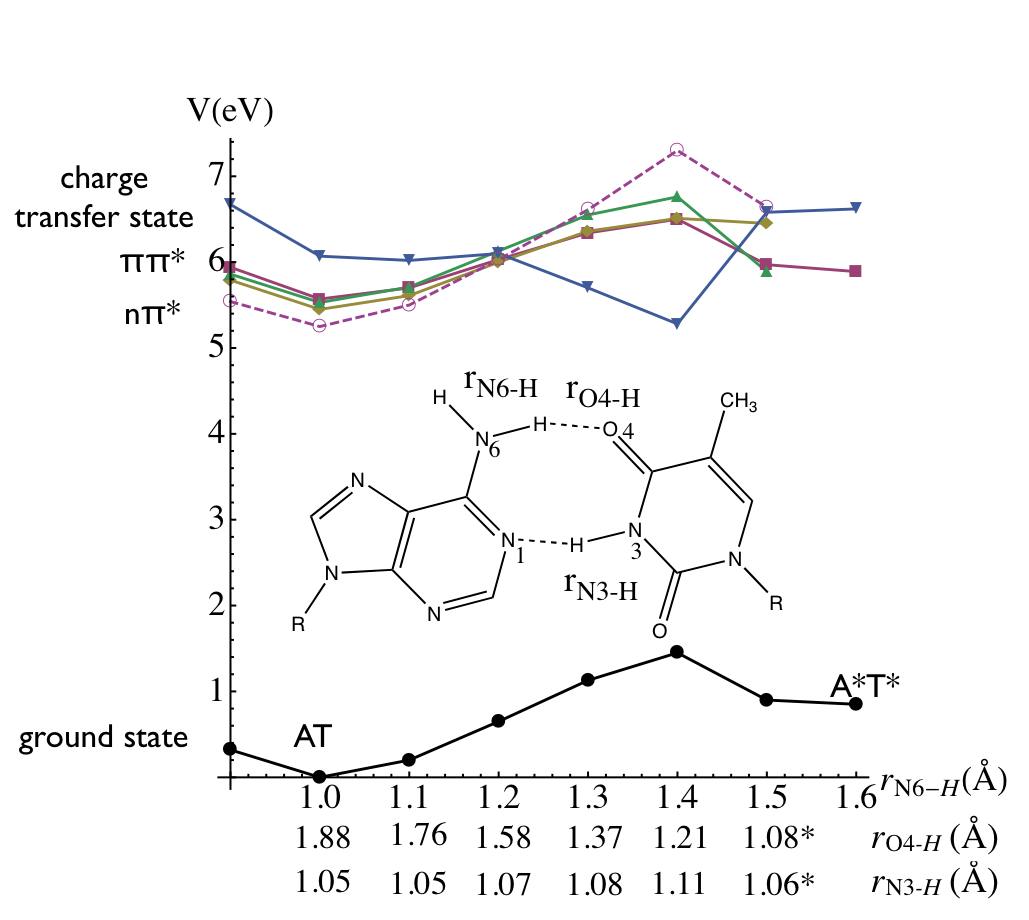}}
\caption{{\em Ab initio} ground and excited state potential curves for a dAdT Watson/Crick  nucleoside pair along the N6(A)--H stretching coordinate.  At each point along the curve, the ground-state geometry was optimized constraining the sugars to their positions in a B-DNA chain.} 
\label{fig1}
\end{figure}

Similarly, by comparing the ultrafast transient absorption signals for isolated bases, single-stranded oligonucleosides and base-paired oligonucleosides, Kohler's group concludes that because the ground state recovery of single and double strand systems are essentially identical and very different from isolated bases.\cite{Pecourt:2000,Pecourt,pecourt:10370,Crespo-Hernandez:2005,Crespo-Hernandez:2006}
These transient absorption on dAdT oligonucleotides support a model whereby the initial local excitation evolves on an ultrafast time scale to form a long-lived species that subsequntly decays to the electronic ground state.  Kohler {\em et al.} assigns the long-lived species to inter-base charge-transfer states in which an electron is transferred from one base to a $\pi$-stacked neighbor to form an intrastrand charge-transfer exciplex. While the precise nature of these long-lived states has been subject to debate, \cite{Markovitsi:2006,Crespo-Hernandez:2006} it is clear from these results that base stacking interactions as opposed to the electron-driven proton-transfer dictates the excited state dynamics in double-stranded DNA. 
 
In contrast to these results, experiments and calculations by Domcke and coworkers\cite{schultz:1765,Sobolewski:2002, Sobolewski:2005vn} on isolated base-pairs and pyridine dimers indicate that upon photoexcitation to a $\pi\pi^*$ state, the transfer of a proton from one molecule to the other carries the system through two conical intersections,  first between an excitonic $\pi\pi^*$ state and $\pi\pi^*$ charge-transfer (CT) state and second between the CT state and the ground-state on a timescale of $\approx$ 60--100  ps in matrix-isolated model base-pair dimer systems depending upon system. Based upon these results, a back and forth proton transfer could account for the intrinsic photo-stability in actual DNA chains. 

Here, we report on our series of theoretical investigations aimed at providing a unified dynamical model for describing the internal photochemical relaxation pathways in a DNA oligomer. \cite{bittner:094909,Czader:2008} Our studies combine accurate and high-level quantum chemical descriptions of the relevant excited states of a Watson-Crick AT nucleoside pair.  Stacking interactions are introduced by computing the exact Coulomb coupling between excited state transition densities on neighboring base-pairs as sampled from a molecular dynamics simulation of a 12-base DNA oligomer of d(A)$\cdot$d(T) in water in its ``B'' form. By including both base-stacking, base-pairing, and proton exchange on an equal footing, we hope to provide an unbiased assessment of the initial photophysical processes. Our results indicate that photoexcited states in DNA chain can be rapidly localized by fluctuations in the protons involved with base pairing and that excited-state proton transfer between paired nucleosides can effectively quench photoexcitations.  Weak coupling between delocalized neutral excitonic states and localized charge-transfer states does account for the enhanced lifetimes of the photoexciations.    

The chemical structure of the dAdT base-pair is shown in the inset of Fig.~\ref{fig1} along with several key bond distances. In agreement with previous calculations both equilibrium structures were found slightly buckled and propeller-twisted. 
\footnote{Unless otherwise noted, all of our calculations were performed using Hartree-Fock (HF) theory with a high-quality basis (cc-pVDZ). Excited states were determined using a  configuration interaction expansion of the wavefunction that included both single and double 
excitation terms. }
In addition, the distance between the two fragments, A and T, is shorter in the ``rare'' tautomer, a feature also calculated by Villani.\cite{Villani_2005} Particularly, the N6-H-O4 hydrogen bridge in the imino/enol-tautomer is shorter by 0.324 \AA~ compared with the corresponding distance in the Watson/Crick structure. The energy profile shown in Fig. \ref{fig1} was constructed by moving the N6 proton of adenine toward thymine and reoptimizing all other bases geometrical parameters. These states are energetically very close together  and resolving them accurately requires both large electronic bases-sets, accurate treatment of electronic-correlations, and the inclusion of virtual excitations. Furthermore the influence of these effects is magnified as the system is pushed towards the two conical intersections encountered as we stretch the $r_{N6-H}$ bond. Finally, we found that there is is considerable interplay between the bases and the sugars that we could not simply replace them with methyl groups. The barrier for the double proton transfer dAdT $\rightarrow$ dA*dT* in the ground state calculated at the HF/cc-pVDZ is rather high, 31.1 kcal/mol, while the barrier for the reverse process dA*dT* $\rightarrow$ dAdT calculated at the same level amounts to 14.9 kcal/mol (Fig.\ref{fig1}). Especially, the latter barrier is significantly higher compared with the values of 0.2 and 0.02 kcal/mol reported by Gorb et al\cite{Gorb_2004} and Guallar et al,\cite{Guallar_1999} respectively, for a AT base pair without sugars. However, reoptimization of the structures with sugars at the MP2/def-SV(P) level lowers the barriers for the forward and reverse processes to 18.2 and 2.0 kcal/mol respectively (data not shown).

\begin{figure}[t]
\center{
\includegraphics[width=1.0\columnwidth]{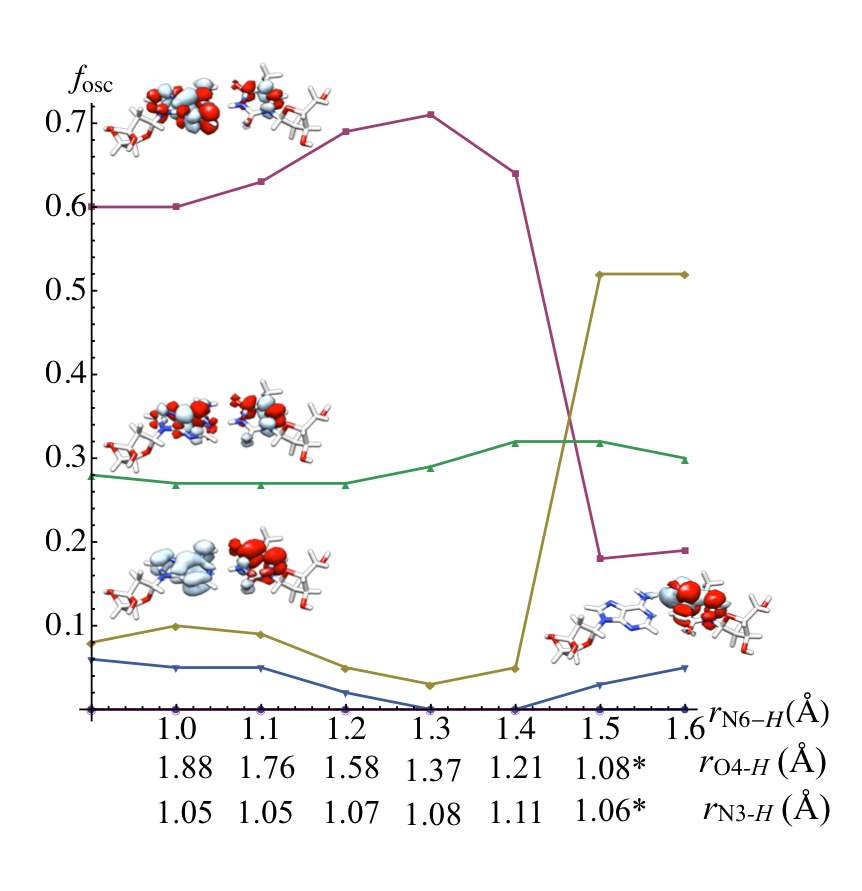}}
\caption{Plot of the oscillator strength corresponding to the $n\pi*$ and $\pi\pi*$ states as a function N6(A)--H bond distance. The difference densities shown next to each trace correspond to the ground state configuration of the WC AT base pair.}
\label{fig2}
\end{figure}

Vertical excitation energies (VEE) of the lowest $n\pi^*$, localy excited   $\pi\pi*$ excitonic states and charge transfer $\pi\pi*$ states of the Watson-Crick base pair calculated as a function of the N6-H distance of adenine at the CIS(D) level are plotted in Fig.~\ref{fig1}. At the ground state equilibrium geometry, the calculated lowest energy excited state corresponds to the $n\pi*$ state with both the $n$ and $\pi*$ orbitals completely localized on thymine. The $\pi\pi*$ exciton states are just about 0.2--0.4 eV above the $n\pi*$ state.  We also determine that  state corresponding to the charge-transfer ($\pi\pi*$ CT) state is approximately 0.6 eV above the $\pi\pi*$ localized exciton states. The vertical excitation energy calculated at CIS(D) level are generally in good agreement with those calculated at the CC2 level for the AT base pair without sugars using similar quality basis set\cite{Perun_2006}. The primary configurations in all three $\pi\pi^*$ exciton states correspond to electron/hole excitations localized on either thymine or adenine bases. The weights of these configurations are by and large similar, below 50\%, producing the  $\pi\pi^*$ excited states delocalized over both bases with none of them entirely localized on just one base. The calculated vertical excitation energy of the $\pi\pi^*$  exciton states increase while the vertical excitation of the $\pi\pi*$ CT state decreases with increasing N6--H (adenine) bond distance up to 1.4 \AA. As the adenine N6--H  bond distance becomes longer this trend is reversed. Furthermore, the character of the excited states also changes. The $\pi\pi*$ exciton states become increasingly localized on just one base either adenine or thymine with the contribution from the corresponding excitation greater than 80\%. Ultimately, this is related to the formation of the tautomer, When the dAdT base pair geometry is reoptimized with the N6--H (adenine) bond distance $r_H = 1.5$ \AA~ the N3 hydrogen of thymine moves over to the other side forming a covalent bond with the adenine N1 to form the imino/enol tautomer.

\begin{figure}[t]
\center{
\includegraphics[width=1.0\columnwidth]{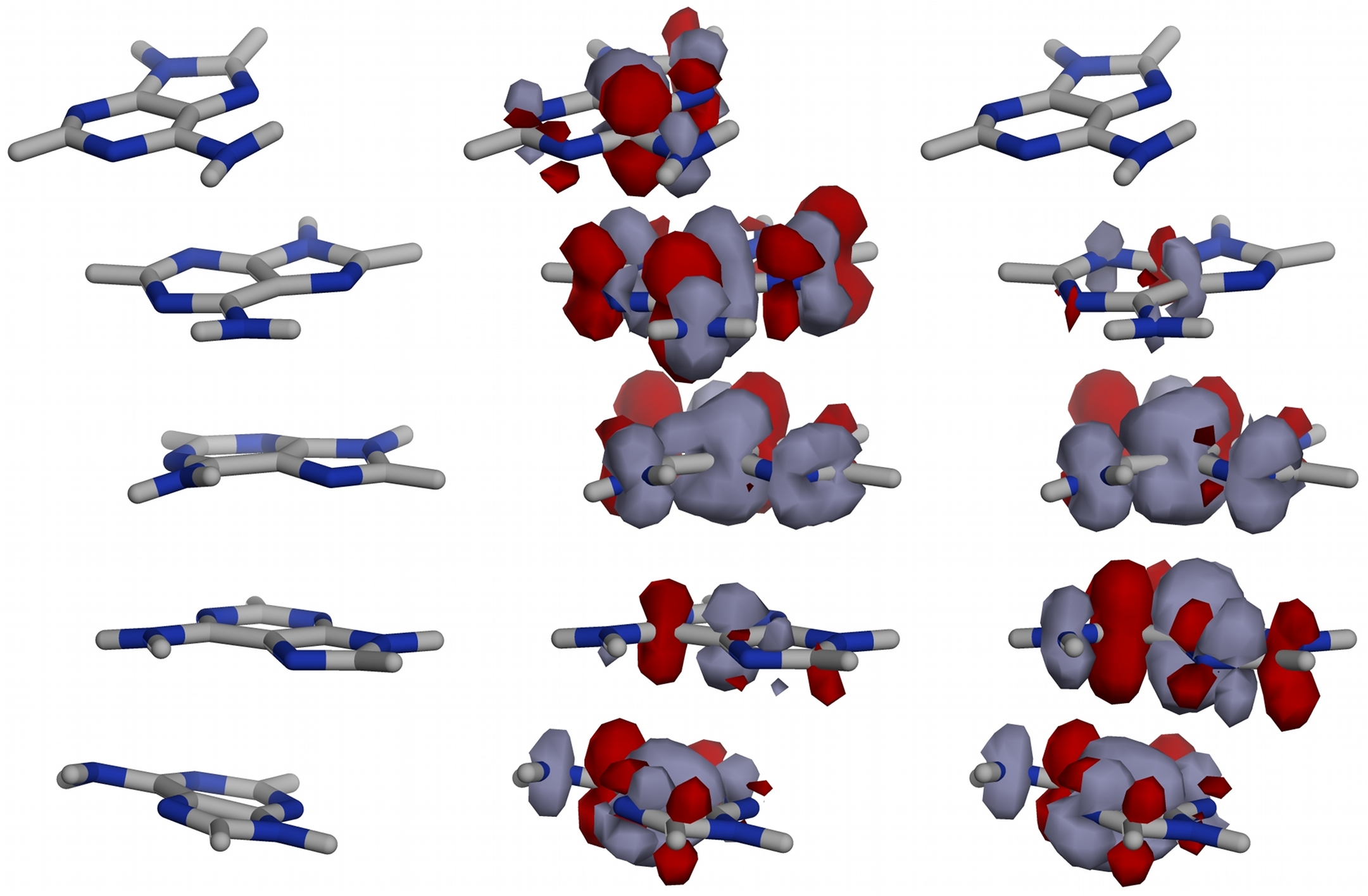}}
\caption{The difference densities of stacked 9H-adenine pentamer (left) calculated at the CIS(D)/cc-pVDZ level corresponding to two different localied $\pi\pi*$ states (middle and right.}
\label{pentamer}
\end{figure}

We also computed the excitation energies for stacked pentamer of 9H-adenine and saw little evidence of charge transfer exciplex states between stacked bases. None of the 20 excited states calculated at the CIS(D)/cc-pVDZ level have a charge transfer character. The lowest energy states can be most adequately classified as $n\pi*$ and $\pi\pi*$ states. Likewise the vertical excitation energies (VEE) of the AT base pair the VEEs of the former states of stacked pentamer are slightly lower than the VEEs of the latter states. The $\pi\pi*$ states with the largest oscillator strength are also the most delocalized, the corresponding molecular orbitals being delocalized over all five adenine bases. On the contrary, the $n\pi*$ states can be localized on a single thymine base. Preliminary CIS(D) calculations on stacked adenosine pentamers based upon MD configurations give little evidence for exciplex formation, although depending upon the instantaneous geometry of the chain, excitons may have a small static dipole due to differing degrees of localization of the occupied and virtual orbitals contributing to the configuration-interaction expansion of the excited state wave functions. The difference densities corresponding to the $\pi\pi*$ states of the pentamer with the largest oscillator strength are shown in Fig.~\ref{pentamer}. These calculations were performed in vacuum, consequently,  it is entirely possible that contributions from solvent polarization, counter-charges, and near by water molecules could stabilize intrastrand exciplex states.  One can argue against exciplex formation in stacked homodimers since the exciton binding energy ($\approx 0.4--0.5$ eV typical for conjugated organic species) is far greater than the difference in either electron affinities or ionization potentials of the stacked pair. \cite{bredas:447,karabunarliev:3988}  Based upon the exciplex stability criteria, exciplex states are only expected for stacked heterodimers (eg. AT, CG, etc...)

For the purpose of developing a minimal model for the excited states of DNA chain we need to make a number of judicious assumptions. The Coulombic coupling  elements between two stacked Watson-Crick AT base pairs were calculated using the transition density cube (TDC) method described in our previous report\cite{Czader:2008}. The transition densities have been calculated for several Watson-Crick AT base pairs molecular geometries extracted from the molecular dynamics simulations. In Fig.~\ref{fig2} we show the transition densities calculated at the CIS/cc-pVDZ level corresponding to two different localied $\pi\pi*$ states superimposed at the neighboring AT base pairs of the (dA)$_{10}\cdot$(dT)$_{10}$. 

\begin{figure}[t]
\begin{center}
\includegraphics[width=0.90\columnwidth]{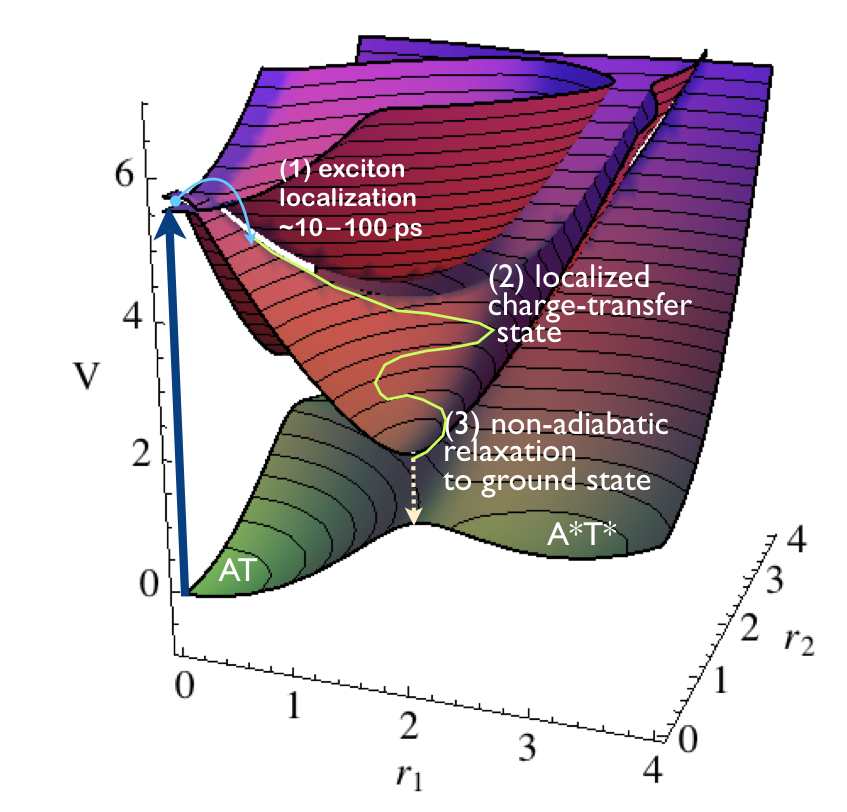}
\end{center}
\caption{The three-dimensional potential energy surface describing the motion of protons between N6(A) and O4(T) and between N3(T) and N1(A) shows two critical points in the ground state. The deeper minimum corresponds to the amine/keto structure of AT and a shallow one to the imine/enol structure (A*T*). Upon absorption of a UV photon (blue arrow) the initaly delocalized excitonic states (1) undergo a rapid localization on  $\approx $10 ps timescale for single bases and $\approx $100 ps timescale for stacked base pairs to form a charge transfer (CT) states (2). The subsequent CT states passing through a conical intersection are carried back to the ground state (3).} 
\label{fig3}
\end{figure}

Our quantum chemical results suggest that a minimal model for the relevant states of a DNA oligomer must consist of a ground state $|g\>$ and localized $\pi\pi*$ excitons $(|e_j\>)$ and local $\pi\pi^*$ charge-transfer states $(|c_j\>)$ for a each Watson-Crick pair.  We assume that only these states are involved in the delocalization and relaxation process following photoexcitation. We assume that the overlap between neighboring states is sufficiently small that $\<e_n|e_{n'}\> = \<c_n | c_{n'}\> = \delta_{nn'}$ and $\<e_n|c_n\> = 0$. The coupling interactions between neighboring excitonic states $|e_j\>$ and $|e_{j\pm1}\>$ ($\lambda$) were determined to be in the range 200--400 cm$^{-1}$. This sets the time scale for coherent exciton transfer between stacked bases to be on the order of 5-10 fs. 

The excitonic couplings are generally much stronger than those between the local excitonic and CT states ($j_2$ = 50 cm$^{-1}$) or between two stacked CT $\pi\pi*$ states ($|c_j\>$ and $|c_{j\pm 1}\>$) (20 cm$^{-1}$) as calculated using the transition density cube. These stacking interaction are highly sensitive to the relative orientation of base pairs and can change by an order of magnitude. However, the absolute magnitude of the interaction is very small compared with the vertical excitation energies of the d(A)$\cdot$d(T) base pair calculated at CIS(D)/cc-pVDZ level. Consequently, if we estimate the energy lowering of  excitonic $\pi\pi*$ states due to the excitonic coupling as twice the nearest-neighbor coupling calculated using TDC method the delocalized exciton is about 800 cm$^{-1}$ lower in energy than a localized exciton.  
Finally we include coupling between the the ground state $|g\>$ and the $\pi\pi*$ states $|e_j\>$ ($j_1$) and the ground state and  the CT  states ($j_3$). We systematically adjusted these terms as to best reproduce the energies of the ground state and the lowest energy $\pi\pi^*$ states of the d(A)$\cdot$d(T) nucleoside from our {\em ab initio} studies. (Fig.\ref{fig1}).  The adiabatic energy surfaces along the proton-transter normal mode coordinate $r_n \propto (\delta_{(A)N6-H} + \delta_{(T)N3-H})$ for a dimer can be obtained from the eigenvalues of the following $5\times 5$ matrix:
\begin{widetext}
\begin{eqnarray}
\left(\begin{array}{ccccc}
E_g & j_1 & j_3 & j_1 & j_3 \\
j_1 & E_{e1} + g_1 r_1/2 & j_2 & \lambda & 0 \\
j_3 & j_2 & E_{c1} + g_2 r_1/2& 0 & 0 \\
j_1 & 0 & 0 & E_{e2}  + g_1 r_2/2 & j_2 \\
j_3 & 0 & 0 & j_2 & E_{c2} + g_2 r_2/2\end{array}\right) +H_{protons}
\end{eqnarray}
\end{widetext}

% more details of potential surface topology
Fig.~\ref{fig3} shows the adiabatic potentials arising from our reduced model. First, the lowest surface is the potential surface for tautomerization in the ground state.  Two minima occur at $(r_1, r_2) = (3,0)$ and $(0,3)$ corresponding to one base pair or the other in the tautomer form. However, both excitonic states are unstable with respect to the proton exchange coordinates.  Once the system has moved away from the origin along one of the proton-transfer coordinates, the electronic states rapidly localize and we are carried towards the conical intersection between the local CT state and the ground state.

Let us assume that the lifetime of the delocalized state is limited by proton transfer between one of the base pairs such that as soon as one proton coordinate cross the XT/CT intersection, the delocalized state collapses to from a localized state. Assuming the usual Condon separation between the nuclear and electronic dynamics, we can write this within the 
non-adiabatic Marcus approximation
\begin{eqnarray}
k_{loc} = \frac{2\pi}{\hbar}|V_{ab}|^2 \frac{1}{\sqrt{4\pi E_r k_BT}}e^{-(\Delta E + E_r)^2/(4E_r k_B T)}
\end{eqnarray}
where $V_{if} = j_2$ is the diabatic coupling.  We can estimate this rate by setting the driving force $\Delta E$ to be the energy difference between the vertical exciton and the ground-state tautomer and the reorganization energy $E_r$ as $E_r = E_{ct} - E_{t}$. This sets the time-scale for interbase electron transfer of $\tau = 1/k_{loc}  = 10.3$ ps.  This gives a lower limit to the exciton lifetime since even  a small error in our values can change this by a factor of 2 to 5. Moreover, for the delocalized case, the coupling matrix element will be at least proportional to the probability for finding the exciton on a given site, $V_{ab} \propto j_2 \rho_{n}$. Thus, for the delocalized case where the exciton is extended over 3--5 bases  we expect $\tau \approx 100$ to 250 ps.  
In summary, this study combined with our previous study of exciton delocalization in B-DNA chains~\cite{Czader:2008} proposes the following mechanism. Following vertical $\pi-\pi^*$ excitation of an adenosine, the exciton rapidly delocalizes between 3--4 neighboring stacked A's on a time-scale given by the exciton-exciton coupling, $\lambda$. The delocalization length is limited by the fact that $\lambda$ is strongly modulated by the structural fluctuations of the DNA chain about its ideal B-DNA form. This initial delocalization occurs on the femtosecond time-scale. Next, these states are unstable with respect to the fluctuations of the stretching motions of the protons involved in the hydrogen-bonding between the Watson/Crick pairs.  This causes a re-localization of the exciton to occur on the time-scale of pico-seconds. Subsequent relaxation to the ground-state occurs on a longer time scale as determined by the conical intersection between the CT and the ground-state.    

Although this study has focused solely on poly(dA)$\cdot$-poly(dT) chains,  the implication for more general sequences is clear. It is entirely possible that such internal photodeactivation mechanisms are responsible for the intrinsic photo-stability of DNA.  Such mechanisms may have provided a necessary evolutionary advantage for early DNA-based life forms.

{\bf Acknowledgments:} 
This work was funded in part by the National Science Foundation,  
and the Robert A. Welch Foundation.

\end{document}